\documentclass[11pt]{article}
\usepackage{hyperref}
\usepackage[pdftex]{graphicx}
\pdfoutput=1
\usepackage{hyperref}
\pdfoutput=1

\usepackage[margin=0.5in]{geometry}
\date{}
\begin{document}
\title{3D Shock-Bubble Interactions at Mach 3}
\author{Babak Hejazialhosseini, Diego Rossinelli and Petros Koumoutsakos \\
Computational Science and Engineering Laboratory, ETH Zurich, Switzerland}
\maketitle
\begin{figure*}[h]
\centering
\includegraphics[width=0.8\textwidth]{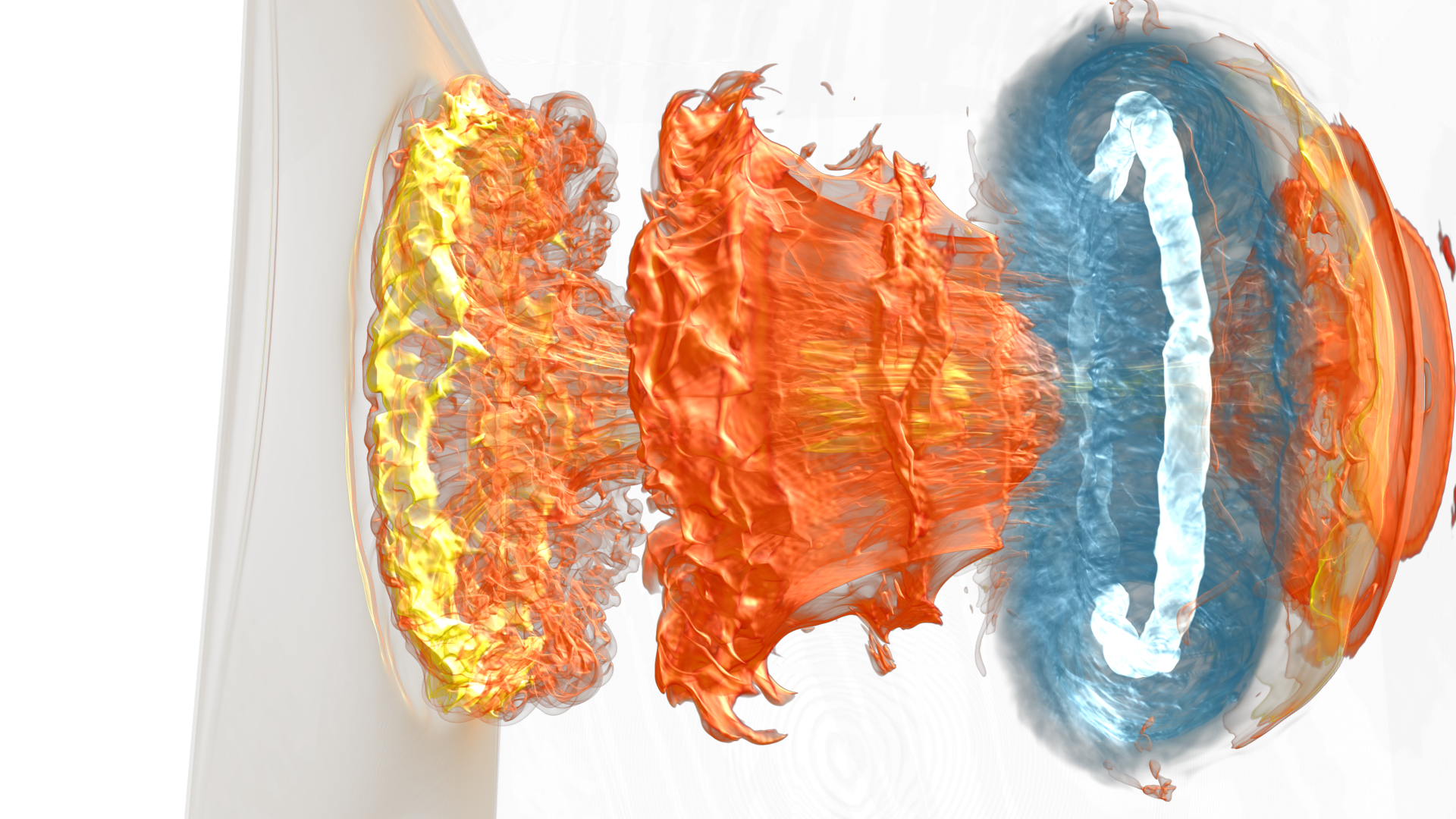}
\caption{Volume-rendered density field of late-time shock-bubble interactions at M=3. Orange/blue denote high/low density values.}
\label{fig:sbi}
\end{figure*}

We present a simulation for the interactions of shockwaves with light spherical density inhomogeneities. Euler equations for two-phase compressible flows are solved in a 3D uniform resolution finite volume based solver using 5th order WENO reconstructions \cite{Jiang:1996} of the primitive quantities \cite{Johnsen:2006}, HLL-type numerical fluxes \cite{Toro:1994} and 3rd order TVD time stepping scheme. In this study, a normal Mach 3 shockwave in air is directed at a helium bubble with an interface Atwood number of -0.76. We employ 4 billion cells on a supercomputing cluster and demonstrate the development of this flow until relatively late times. Shock passage compresses the bubble and deposits baroclinic vorticity on the interface. Initial distribution of the vorticity and compressions lead to the formation of an air jet, interface roll-ups and the formation of a long lasting vortical core (see Fig. \ref{fig:sbi}, the white core). Compressed upstream of the bubble turns into a mixing zone and as the vortex ring distances from this mixing zone, a plume-shaped region is formed and sustained. Close observations have been reported in previous experimental works \cite{Ranjan:2008}. The visualization is presented in a \href{puthelink}{fluid dynamics video}.

\end{document}